# A note on proper affine vector fields in special non static axially symmetric Lorentzian manifolds

Ghulam Shabbir[1], K S Mahomed[2] and R J Moitsheki[2]

[1]Faculty of Engineering Sciences, GIK Institute of Engineering Sciences and Technology, Topi, Swabi, KPK, Pakistan.

Email: shabbir@giki.edu.pk

[2]School of Computer Science and Applied Mathematics, University of the Witwatersrand, Johannesburg, Wits 2050, South Africa

## Abstract

A study of proper affine vector fields in non static axially symmetric space-times is given by using holonomy and decomposability, the rank of the $6\times6$ Riemann matrix and direct integration techniques. It is shown that the special class of the above space-times admits proper affine vector fields.

## 1. Introduction

The role of symmetry is vital in order to understand the Universe. The general theory of relativity is well described mathematically by Einstein's field equations, in which one side tells us the geometry of the space-times and the other side gives its physics. Einstein's field equations are highly nonlinear in most of the cases it is impossible to find the exact solutions. Different symmetries restrictions are imposed to find the solutions of these highly nonlinear equations. The importance of the study of symmetries is quite clear in general relativity because the laws of conservation in the space-time can be studied and well understood with the help of these symmetry restrictions [1]. The geometrical features of the space-time can be understood and a wide range of physical information about the space-time can be obtained through symmetries. For example in general relativity self-similarity solutions are broadly used for star formation, cosmological perturbations, primordial black holes, gravitational collapse, cosmic censorship and cosmological voids [2]. Keeping in mind the importance and interest on symmetries, we wish to find the existence of proper affine vector fields in the special non-static axially symmetric space-times. Different approaches were adopted to study affine vector fields [3-12]. Here, we will use holonomy and decomposability, the rank of the $6\times 6$ Rieman matrix and direct integration techinques to study the proper affine vector fields. Throughout $M$ represents a four dimensional, connected, Hausdorff space-time manifold with Lorentz metric $g$ of signature $(-,+,+,+)$. The curvature tensor associated with $g_{ab}$, through the Levi-Civita connection, is denoted in component form by $R^{a}{}_{bcd}$, and the Ricci tensor components are $R_{ab} = R^{c}{}_{acb}$. The usual covariant, partial and Lie derivatives are denoted by a semicolon, a comma and the symbol $L$, respectively. Round and square brackets denote the usual symmetrization and skew-symmetrization, respectively. Here, $M$ is assumed to be non-flat in the sense that the curvature tensor does not vanish over any non-empty open subset of $M$.

A vector field $X$ on $M$ is called an affine vector field if it satisfies

$$X_{a;bc} = R_{abcd} X^{d} \tag{1}$$

or equivalently,

$$X_{a,bc} - \Gamma^{f}_{ac} X_{f,b} - \Gamma^{f}_{bc} X_{a,f} - \Gamma^{e}_{ab} X_{e,c} + \Gamma^{e}_{ab}\Gamma^{f}_{ec} X_{f} - (\Gamma^{e}_{ab})_{,c} X_{e} - \Gamma^{f}_{ab}\Gamma^{e}_{cf} X_{e} + \Gamma^{e}_{fb}\Gamma^{f}_{ca} X_{e} + \Gamma^{e}_{af}\Gamma^{f}_{bc} X_{e} = R_{abcd} X^{d}.$$

If one decomposes $X_{a;b}$ on $M$ into its symmetric and skew-symmetric parts

$$X_{a;b} = \frac{1}{2} h_{ab} + F_{ab} \qquad (h_{ab} = h_{ba}, \;\; F_{ab} = -F_{ba}), \tag{2}$$

then equation (1) is equivalent to

$$(i)\; h_{ab;c} = 0 \quad (ii)\; F_{ab;c} = R_{abcd} X^{d} \quad (iii)\; F_{ab;c} X^{c} = 0. \tag{3}$$

The proof of the above equation (1) implies (3) or equations (3) implies (1) can be found in [4,5]. If $h_{ab} = 2cg_{ab}, c \in R,$ then the vector field $X$ is called homothetic (and *Killing* if $c = 0$). The vector field $X$ is said to be proper affine if it is not homothetic vector field and also $X$ is said to be proper homothetic vector field if it is not a Killing vector field on $M$ [4]. We define the subspace $S_p$ of the tangent space $T_pM$ to $M$ at $p$ as those $k \in T_pM$ satisfying

$$R_{abcd}k^d = 0. \tag{4}$$

## 2. Affine Vector Fields

Suppose that $M$ is a simple connected space-time. Then the holonomy group of $M$ is a connected Lie subgroup of the idenity component of the Lorentz group and is thus characterized by its subalgebra in the Lorentz algebra. These have been labeled into fifteen types $R_1 - R_{15}$ [12]. It follows from [4] that the only space-times which could admit proper affine vector fields are those which admit nowhere zero covariantly constant second order symmetric tensor field $h_{ab}$. This forces the holonomy type to be either $R_2$, $R_3$, $R_4$, $R_6$, $R_7$, $R_8$, $R_{10}$, $R_{11}$ or $R_{13}$ [4]. A study of the affine vector fields for the above holonomy types can be found in [4]. It follows from [4] that the rank of the $6\times 6$ Riemann matrix of the above space-times which have holonomy type $R_2$, $R_3$, $R_4$, $R_6$, $R_7$, $R_8$, $R_{10}$, $R_{11}$ or $R_{13}$ is at the most three. Hence for studying affine vector fields we are interested in those cases when the rank of the $6\times 6$ Riemann matrix is less than or equal to three.

## 3. Main Results

Consider a special non static axially symmetric space-time in the usual coordinate system $(t,r,\theta,\phi)$ (labeled by $(x^0,x^1,x^2,x^3)$, respectively) with line element [13]

$$ds^2 = -e^{A(t,r,\theta)}dt^2 + e^{B(t,r,\theta)}(dr^2 + d\theta^2 + d\phi^2) \tag{5}$$

The above space-time admits only one Killing vector field which is $\frac{\partial}{\partial\phi}$. The non-zero independent components of the Riemann tensor are [14]

$$R_{0101} = \frac{1}{4}\begin{bmatrix} e^{A(t,r,\theta)}(A_r^2(t,r,\theta) + 2A_{rr}(t,r,\theta) - A_r(t,r,\theta)B_r(t,r,\theta) + A_\theta(t,r,\theta)B_\theta(t,r,\theta)) \\ -e^{B(t,r,\theta)}(B_t^2(t,r,\theta) + 2B_{tt}(t,r,\theta) - A_t(t,r,\theta)B_t(t,r,\theta)) \end{bmatrix} \equiv \alpha_1,$$

$$R_{0102} = \frac{1}{4}e^{A(t,r,\theta)}\begin{bmatrix} A_r(t,r,\theta)A_\theta(t,r,\theta) + 2A_{r\theta}(t,r,\theta) \\ -A_r(t,r,\theta)B_\theta(t,r,\theta) - A_\theta(t,r,\theta)B_r(t,r,\theta) \end{bmatrix} \equiv \alpha_2,$$

$$R_{0112} = -R_{0323} = \frac{1}{4}e^{B(t,r,\theta)}\left[2B_{t\theta}(t,r,\theta) - A_\theta(t,r,\theta)B_t(t,r,\theta)\right] \equiv \alpha_3,$$

$$R_{0202}=\frac{1}{4}\begin{bmatrix} e^{A(t,r,\theta)}(A_\theta^2(t,r,\theta)+2A_{\theta\theta}(t,r,\theta)-A_\theta(t,r,\theta)B_\theta(t,r,\theta)+A_r(t,r,\theta)B_r(t,r,\theta)) \\ -e^{B(t,r,\theta)}(B_t^2(t,r,\theta)+2B_{tt}(t,r,\theta)-A_t(t,r,\theta)B_t(t,r,\theta)) \end{bmatrix}\equiv\alpha_4,$$

$$R_{0212}=R_{0313}=-\frac{1}{4}e^{B(t,r,\theta)}\left[2B_{tr}(t,r,\theta)-A_r(t,r,\theta)B_t(t,r,\theta)\right]\equiv\alpha_5,$$

$$R_{0303}=\frac{1}{4}\begin{bmatrix} e^{A(t,r,\theta)}(A_r(t,r,\theta)B_r(t,r,\theta)+A_\theta(t,r,\theta)B_\theta(t,r,\theta)) \\ -e^{B(t,r,\theta)}(B_t^2(t,r,\theta)+2B_{tt}(t,r,\theta)-A_t(t,r,\theta)B_t(t,r,\theta)) \end{bmatrix}\equiv\alpha_6,$$

$$R_{1212}=-\frac{1}{4}e^{B(t,r,\theta)-A(t,r,\theta)}\left[2(B_{rr}(t,r,\theta)+B_{\theta\theta}(t,r,\theta))e^{A(t,r,\theta)}-e^{B(t,r,\theta)}B_t^{\ 2}(t,r,\theta)\right]\equiv\alpha_7,$$

$$R_{1313}=-\frac{1}{4}e^{B(t,r,\theta)-A(t,r,\theta)}\left[(B_\theta^{\ 2}(t,r,\theta)+B_{rr}(t,r,\theta))e^{A(t,r,\theta)}-e^{B(t,r,\theta)}B_t^{\ 2}(t,r,\theta)\right]\equiv\alpha_8,$$

$$R_{1323}=-\frac{1}{4}e^{B(t,r,\theta)}\left[2B_{r\theta}(t,r,\theta)-B_r(t,r,\theta)B_\theta(t,r,\theta)\right]\equiv\alpha_9,$$

$$R_{2323}=-\frac{1}{4}e^{B(t,r,\theta)-A(t,r,\theta)}\left[(B_r^{\ 2}(t,r,\theta)+2B_{\theta\theta}(t,r,\theta))e^{A(t,r,\theta)}-e^{B(t,r,\theta)}B_t^{\ 2}(t,r,\theta)\right]\equiv\alpha_{10}.$$

We write the curvature tensor with components $R_{abcd}$ at point $p$ of the manifold as the $6\times 6$ symmetric matrix

$$R_{abcd}=\begin{bmatrix} \alpha_1 & \alpha_2 & 0 & \alpha_3 & 0 & 0 \\ \alpha_2 & \alpha_4 & 0 & \alpha_5 & 0 & 0 \\ 0 & 0 & \alpha_6 & 0 & \alpha_5 & -\alpha_3 \\ \alpha_3 & \alpha_5 & 0 & \alpha_7 & 0 & 0 \\ 0 & 0 & \alpha_5 & 0 & \alpha_8 & \alpha_9 \\ 0 & 0 & -\alpha_3 & 0 & \alpha_9 & \alpha_{10} \end{bmatrix} \tag{6}$$

As mentioned in section 2, the space-times which can admit proper affine vector fields have holonomy type $R_2$, $R_3$, $R_4$, $R_6$, $R_7$, $R_8$, $R_{10}$, $R_{11}$ or $R_{13}$ and the rank of the $6\times 6$ Riemann matrix is at the most three. Therefore we are only interested in those cases when the rank of the $6\times 6$ Riemann matrix is less than or equal to three. It follows from [14] that there exist the following possibilities when the rank of the $6\times 6$ Riemann matrix is three or less. These are:

(P1) Rank=3, $A_t(t,r,\theta)\neq 0$, $A_r(t,r,\theta)=0$, $A_\theta(t,r,\theta)\neq 0$, $B_t(t,r,\theta)=0$, $B_r(t,r,\theta)\neq 0$, $B_\theta(t,r,\theta)=0$, $B_{rr}(t,r,\theta)=0$ and $A_\theta^2(t,r,\theta)+2A_{\theta\theta}(t,r,\theta)=0$.

(P2) Rank=3, $A_t(t,r,\theta)\neq 0$, $A_r(t,r,\theta)=0$, $A_\theta(t,r,\theta)\neq 0$, $B_t(t,r,\theta)=0$, $B_r(t,r,\theta)\neq 0$, $B_\theta(t,r,\theta)=0$, $B_{rr}(t,r,\theta)=0$ and $A_\theta^2(t,r,\theta)+2A_{\theta\theta}(t,r,\theta)\neq 0$.

(P3) Rank=3, $A_t(t,r,\theta)\neq 0$, $A_r(t,r,\theta)\neq 0$, $A_\theta(t,r,\theta)=0$, $B_t(t,r,\theta)=0$, $B_r(t,r,\theta)=0$, $B_\theta(t,r,\theta)\neq 0$, $B_{\theta\theta}(t,r,\theta)=0$ and $A_r^2(t,r,\theta)+2A_{rr}(t,r,\theta)=0$.

(P4) Rank=3, $A_t(t,r,\theta)\neq 0$, $A_r(t,r,\theta)\neq 0$, $A_\theta(t,r,\theta)=0$, $B_t(t,r,\theta)=0$, $B_r(t,r,\theta)=0$, $B_\theta(t,r,\theta)\neq 0$, $B_{\theta\theta}(t,r,\theta)=0$ and $A_r^2(t,r,\theta)+2A_{rr}(t,r,\theta)\neq 0$.

(Q1) Rank=3, $A_t(t,r,\theta)\neq 0$, $A_r(t,r,\theta)=0$, $A_\theta(t,r,\theta)=0$, $B_t(t,r,\theta)=0$, $B_r(t,r,\theta)\neq 0$, $B_\theta(t,r,\theta)\neq 0$, $B_r(t,r,\theta)B_\theta(t,r,\theta)-2B_{r\theta}(t,r,\theta)=0$ and $B_{rr}(t,r,\theta)+B_{\theta\theta}(t,r,\theta)\neq 0$.

(Q2) Rank=3, $A_t(t,r,\theta)\neq 0$, $A_r(t,r,\theta)=0$, $A_\theta(t,r,\theta)=0$, $B_t(t,r,\theta)=0$, $B_r(t,r,\theta)\neq 0$, $B_\theta(t,r,\theta)\neq 0$, $B_r(t,r,\theta)B_\theta(t,r,\theta)-2B_{r\theta}(t,r,\theta)\neq 0$ and $B_{rr}(t,r,\theta)+B_{\theta\theta}(t,r,\theta)\neq 0$.

(Q3) Rank=3, $A_t(t,r,\theta)\neq 0$, $A_r(t,r,\theta)=0$, $A_\theta(t,r,\theta)=0$, $B_t(t,r,\theta)=0$, $B_r(t,r,\theta)=0$, $B_\theta(t,r,\theta)\neq 0$ and $B_{\theta\theta}(t,r,\theta)\neq 0$.

(Q4) Rank=3, $A_t(t,r,\theta)\neq 0$, $A_r(t,r,\theta)=0$, $A_\theta(t,r,\theta)=0$, $B_t(t,r,\theta)=0$, $B_\theta(t,r,\theta)=0$, $B_r(t,r,\theta)\neq 0$ and $B_{rr}(t,r,\theta)\neq 0$.

(Q5) Rank=2, $A_t(t,r,\theta)\neq 0$, $A_r(t,r,\theta)=0$, $A_\theta(t,r,\theta)=0$, $B_t(t,r,\theta)=0$, $B_r(t,r,\theta)\neq 0$, $B_\theta(t,r,\theta)\neq 0$, $B_{rr}(t,r,\theta)=0$, $B_{\theta\theta}(t,r,\theta)=0$ and $B_r(t,r,\theta)B_\theta(t,r,\theta)-2B_{r\theta}(t,r,\theta)\neq 0$.

(Q6) Rank=2, $A_t(t,r,\theta)\neq 0$, $A_r(t,r,\theta)\neq 0$, $A_\theta(t,r,\theta)\neq 0$, $B_t(t,r,\theta)=0$, $B_r(t,r,\theta)=0$, $B_\theta(t,r,\theta)=0$, $A_r^2(t,r,\theta)+2A_{rr}(t,r,\theta)\neq 0$, $A_\theta^2(t,r,\theta)+2A_{\theta\theta}(t,r,\theta)\neq 0$ and $A_r(t,r,\theta)A_\theta(t,r,\theta)+2A_{r\theta}(t,r,\theta)\neq 0$.

(Q7) Rank=2, $A_t(t,r,\theta)\neq 0$, $A_r(t,r,\theta)\neq 0$, $A_\theta(t,r,\theta)\neq 0$, $B_t(t,r,\theta)=0$, $B_r(t,r,\theta)=0$, $B_\theta(t,r,\theta)=0$, $A_r^2(t,r,\theta)+2A_{rr}(t,r,\theta)\neq 0$, $A_\theta^2(t,r,\theta)+2A_{\theta\theta}(t,r,\theta)\neq 0$ and $A_r(t,r,\theta)A_\theta(t,r,\theta)+2A_{r\theta}(t,r,\theta)=0$.

(Q8) Rank=2, $A_t(t,r,\theta)\neq 0$, $A_r(t,r,\theta)\neq 0$, $A_\theta(t,r,\theta)\neq 0$, $B_t(t,r,\theta)=0$, $B_r(t,r,\theta)=0$, $B_\theta(t,r,\theta)=0$, $A_r^2(t,r,\theta)+2A_{rr}(t,r,\theta)=0$, $A_\theta^2(t,r,\theta)+2A_{\theta\theta}(t,r,\theta)\neq 0$ and $A_r(t,r,\theta)A_\theta(t,r,\theta)+2A_{r\theta}(t,r,\theta)\neq 0$.

(Q9) Rank=2, $A_t(t,r,\theta)\neq 0$, $A_r(t,r,\theta)\neq 0$, $A_\theta(t,r,\theta)\neq 0$, $B_t(t,r,\theta)=0$, $B_r(t,r,\theta)=0$, $B_\theta(t,r,\theta)=0$, $A_r^2(t,r,\theta)+2A_{rr}(t,r,\theta)\neq 0$, $A_\theta^2(t,r,\theta)+2A_{\theta\theta}(t,r,\theta)=0$ and $A_r(t,r,\theta)A_\theta(t,r,\theta)+2A_{r\theta}(t,r,\theta)\neq 0$.

(Q10) Rank=2, $A_t(t,r,\theta)\neq 0$, $A_r(t,r,\theta)\neq 0$, $A_\theta(t,r,\theta)\neq 0$, $B_t(t,r,\theta)=0$, $B_r(t,r,\theta)=0$, $A_\theta^2(t,r,\theta)+2A_{\theta\theta}(t,r,\theta)=0$, $A_r^2(t,r,\theta)+2A_{rr}(t,r,\theta)=0$, $B_\theta(t,r,\theta)=0$ and $A_r(t,r,\theta)A_\theta(t,r,\theta)+2A_{r\theta}(t,r,\theta)\neq 0$.

(Q11) Rank=3, $A_t(t,r,\theta)\neq 0$, $A_r(t,r,\theta)=0$, $A_\theta(t,r,\theta)=0$, $B_t(t,r,\theta)\neq 0$, $B_t^2(t,r,\theta)+2B_{tt}(t,r,\theta)-A_t(t,r,\theta)B_t(t,r,\theta)=0$, $B_\theta(t,r,\theta)=0$ and $B_r(t,r,\theta)=0$.

(Q12) Rank=3, $A_t(t,r,\theta)\neq 0$, $A_r(t,r,\theta)=0$, $A_\theta(t,r,\theta)\neq 0$, $B_t(t,r,\theta)=0$, $B_r(t,r,\theta)=0$, $B_\theta(t,r,\theta)\neq 0$, $B_{\theta\theta}(t,r,\theta)=0$ and $A_\theta^2(t,r,\theta)+2A_{\theta\theta}(t,r,\theta)-A_\theta(t,r,\theta)B_\theta(t,r,\theta)=0$.

(Q13) Rank=3, $A_t(t,r,\theta)\neq 0$, $A_r(t,r,\theta)\neq 0$, $A_\theta(t,r,\theta)\neq 0$, $B_t(t,r,\theta)=0$, $B_r(t,r,\theta)=0$, $A_r(t,r,\theta)A_\theta(t,r,\theta)+2A_{r\theta}(t,r,\theta)-A_r(t,r,\theta)B_\theta(t,r,\theta)=0$, $B_\theta(t,r,\theta)\neq 0$, $B_{\theta\theta}(t,r,\theta)=0$ and $A_\theta^2(t,r,\theta)+2A_{\theta\theta}(t,r,\theta)-A_\theta(t,r,\theta)B_\theta(t,r,\theta)=0$.

(Q14) Rank=2, $A_t(t,r,\theta)\neq 0$, $A_r(t,r,\theta)\neq 0$, $A_\theta(t,r,\theta)=0$, $B_t(t,r,\theta)=0$, $B_r(t,r,\theta)\neq 0$, $B_{rr}(t,r,\theta)=0$, $A_r^2(t,r,\theta)+2A_{rr}(t,r,\theta)-A_r(t,r,\theta)B_r(t,r,\theta)=0$ and $B_\theta(t,r,\theta)=0$.

(R1) Rank=1, $A_t(t,r,\theta)\neq 0$, $A_r(t,r,\theta)\neq 0$, $A_\theta(t,r,\theta)=0$, $B_t(t,r,\theta)=0$, $B_r(t,r,\theta)=0$, $B_\theta(t,r,\theta)=0$ and $A_r^2(t,r,\theta)+2A_{rr}(t,r,\theta)\neq 0$.

(R2) Rank=1, $A_t(t,r,\theta)\neq 0$, $A_r(t,r,\theta)=0$, $A_\theta(t,r,\theta)\neq 0$, $B_t(t,r,\theta)=0$, $B_r(t,r,\theta)=0$, $B_\theta(t,r,\theta)=0$ and $A_\theta^2(t,r,\theta)+2A_{\theta\theta}(t,r,\theta)\neq 0$.

(R3) Rank=1, $A_t(t,r,\theta)\neq 0$, $A_r(t,r,\theta)=0$, $A_\theta(t,r,\theta)=0$, $B_t(t,r,\theta)=0$, $B_r(t,r,\theta)=0$, $B_\theta(t,r,\theta)\neq 0$ and $B_{\theta\theta}(t,r,\theta)=0$.

(R4) Rank=1, $A_t(t,r,\theta)\neq 0$, $A_r(t,r,\theta)=0$, $A_\theta(t,r,\theta)=0$, $B_t(t,r,\theta)=0$, $B_r(t,r,\theta)\neq 0$, $B_\theta(t,r,\theta)=0$ and $B_{rr}(t,r,\theta)=0$.

(R5) Rank=1, $A_t(t,r,\theta)\neq 0$, $B_t(t,r,\theta)=0$, $B_\theta(t,r,\theta)=0$ $B_r(t,r,\theta)=0$, $A_r^2(t,r,\theta)+2A_{rr}(t,r,\theta)\neq 0$, $A_\theta(t,r,\theta)\neq 0$, $A_\theta^2(t,r,\theta)+2A_{\theta\theta}(t,r,\theta)=0$, $A_r(t,r,\theta)\neq 0$ and $A_r(t,r,\theta)A_\theta(t,r,\theta)+2A_{r\theta}(t,r,\theta)=0$.

(R6) Rank=1, $A_t(t,r,\theta)\neq 0$, $B_t(t,r,\theta)=0$, $B_\theta(t,r,\theta)=0$ $B_r(t,r,\theta)=0$, $A_\theta(t,r,\theta)\neq 0$, $A_r^2(t,r,\theta)+2A_{rr}(t,r,\theta)=0$, $A_\theta^2(t,r,\theta)+2A_{\theta\theta}(t,r,\theta)\neq 0$, $A_r(t,r,\theta)\neq 0$ and $A_r(t,r,\theta)A_\theta(t,r,\theta)+2A_{r\theta}(t,r,\theta)=0$.

We will discuss each case in turn.

**Case P1**

In this case $A_t(t,r,\theta)\neq 0$, $A_r(t,r,\theta)=0$, $A_\theta(t,r,\theta)\neq 0$, $B_t(t,r,\theta)=0$, $B_r(t,r,\theta)\neq 0$, $B_\theta(t,r,\theta)=0$, $B_{rr}(t,r,\theta)=0$ and $A_\theta^2(t,r,\theta)+2A_{\theta\theta}(t,r,\theta)=0$. The above equations imply that $A(t,\theta)=\ln(D_1(t)\theta+D_2(t))^2$ and $B(r)=ar+b$, where $a,b\in R(a\neq 0)$ and $D_1(t)$ is nowhere zero function of integration and $D_2(t)$ is a function of integration. The rank of the $6\times 6$ Riemann matrix is three and there exists no non trivial solution of equation (4). Substituting the above information in equation (5), the line element takes the form

$$ds^2=-(D_1(t)\theta+D_2(t))^2dt^2+e^{(ar+b)}(dr^2+d\theta^2+d\phi^2). \qquad (7)$$

Substituting the above information into the affine equations, one finds the affine vector field in this case is

$$X=(0,0,0,c_1), \qquad (8)$$

where $c_1\in R$. Here, the affine vector field is a Killing vector field which is $\frac{\partial}{\partial\phi}$. Cases (P2) to (P4) are exactly the same.

**Case Q1**

In this case we have $A_t(t,r,\theta)\neq 0$, $A_r(t,r,\theta)=0$, $A_\theta(t,r,\theta)=0$, $B_t(t,r,\theta)=0$, $B_r(t,r,\theta)\neq 0$, $B_r(t,r,\theta)B_\theta(t,r,\theta)-2B_{r\theta}(t,r,\theta)=0$, $B_{rr}(t,r,\theta)+B_{\theta\theta}(t,r,\theta)\neq 0$, $B_\theta(t,r,\theta)\neq 0$ and the rank of the $6\times 6$ Riemann matrix is three. Here, there exists a unique (up to scaling) nowhere zero vector field $t_a = t_{,a}$ such that $t_{a;b}=0.$ From the Ricci identity $R_{abcd}t^d = 0.$ The above constraints give $B(r,\theta)=\ln(p(r)+q(\theta))^{-2}$ and $A=A(t),$ where $p(r)$ and $q(\theta)$ are nowhere zero functions of integration. Substituting the above information in equation (5) and after a suitable rescaling of $t,$ the line element can written in the form

$$ds^2 = -dt^2 + (p(r)+q(\theta))^{-2}(dr^2 + d\theta^2 + d\phi^2). \qquad (9)$$

The above space-time is clearly 1+3 decomposable. Affine vector fields in this case [4] are

$$X = (c_4 t + c_5)\frac{\partial}{\partial t} + X', \qquad (10)$$

where $c_4, c_5 \in R$ and $X'$ is a homothetic vector field in the induced geometry on each of the three-dimensional submanifolds of constant $t.$ The induced metric $g_{\alpha\beta}$ (where $\alpha,\ \beta = 1,\ 2,\ 3$) with non-zero components are given by

$$g_{11} = g_{22} = g_{33} = (p(r)+q(\theta))^{-2} \qquad (11)$$

A vector field $X'$ is called a homothetic vector field if it satisfies

$$L_{X'}\, g_{\alpha\beta} = 2c\, g_{\alpha\beta}, \qquad c\in R. \qquad (13)$$

One can expand the above equation (13) by using (12) to deduce

$$p_r(r)X^1 + q_\theta(\theta)X^2 + (p(r)+q(\theta))X^1{}_{,1} = c(p(r)+q(\theta)), \qquad (14)$$

$$X^1{}_{,2} + X^2{}_{,1} = 0, \qquad (15)$$

$$p_r(r)X^1 + q_\theta(\theta)X^2 + (p(r)+q(\theta))X^2{}_{,2} = c(p(r)+q(\theta)), \qquad (16)$$

$$X^1{}_{,3} + X^3{}_{,1} = 0, \qquad (17)$$

$$X^2{}_{,3} + X^3{}_{,2} = 0, \qquad (18)$$

$$p_r(r)X^1 + q_\theta(\theta)X^2 + (p(r)+q(\theta))X^3{}_{,3} = c(p(r)+q(\theta)). \qquad (19)$$

Solution of the above equations from (14) to (19) is

$$X^1 = \frac{r}{1+k}c + c_1, \qquad X^2 = \frac{\theta}{1+k}c + c_2, \qquad X^3 = \frac{\phi}{1+k}c + c_3, \qquad (20)$$

where $p(r)=\frac{1+k}{k\,c}\left(\frac{r}{1+k}c+c_1\right)^k$, $q(\theta)=\frac{1+k}{k\,c}\left(\frac{\theta}{1+k}c+c_2\right)^k$ and $c_1,c_2,c_3,k\in R($ $k\neq 0,-1).$ The sub-case when $k=0$ or $k=-1$ will be discussed later. In this case the induced geometry on each of the three-dimensional submanifolds of constant $t$ admit proper homothetic vector field. Affine vector fields in this case are given by use of equations (20) in (11)

$$X^0=c_4t+c_5,\ X^1=\frac{r}{1+k}c+c_1, X^2=\frac{\theta}{1+k}c+c_2, X^3=\frac{\phi}{1+k}c+c_3. \tag{21}$$

One can write the above equation (21) after subtracting homothetic vector fields as

$$X=(c_4t,0,0,0). \tag{22}$$

Clearly in this case the above space-times (9) admit proper affine vector fields. A proper affine vector fields in this case is $tt_a$.

Now consider the sub-case when $k=-1.$ In this case the homothetic vector field is a Killing vector field which is

$$X^1=X^2=0,\ X^3=c_4, \tag{23}$$

where $c_4\in R.$ Proper affine vector fields in this case are given in (22).

Now consider the sub-case when $k=0.$ In this case the induced geometry on each of the three-dimensional submanifolds of constant $t$ admit proper homothetic vector field which is

$$X^1=rc+c_1,\quad X^2=\theta c+c_2,\qquad X^3=\phi c+c_3, \tag{24}$$

where $p(r)=\ln(r\,c+c_1)^{\frac{1}{c}}$, $q(\theta)=\ln(\theta\,c+c_2)^{\frac{1}{c}}$ and $c_1,c_2,c_3,c\in R(c\neq 0).$ Affine vector fields in this case are given by the use of equation (24) in (10) as

$$X^0=\mathrm{c}_4t+c_5,\ X^1=rc+c_1,\ X^2=\theta c+c_2,\ X^3=\phi c+c_3. \tag{25}$$

A proper affine vector field in this case is $tt_a$. Cases (Q2) to (Q10) are precisely the same.

**Case Q11**

In this case $A_t(t,r,\theta)\neq 0$, $A_r(t,r,\theta)=0$, $A_\theta(t,r,\theta)=0$, $B_t(t,r,\theta)\neq 0$, $B_t^2(t,r,\theta)+2B_{tt}(t,r,\theta)-A_t(t,r,\theta)B_t(t,r,\theta)=0$, $B_\theta(t,r,\theta)=0$, $B_r(t,r,\theta)=0$ and the rank of the $6\times 6$ Riemann matrix is three. Here, there exists a unique (up to multiple) $t_a = t_{,a}$ solution of equation (4) but $t_a$ is not covariantly constant. From the above constraints, we have $A = A(t)$ and $B(t) = \ln(a\int e^{\frac{A(t)}{2}} dt + b)^2$, where $a,b \in R (a \neq 0)$. The line element after a suitable rescaling of $t$ can be written as

$$ds^2 = -dt^2 + (at+b)^2 (dr^2 + d\theta^2 + d\phi^2). \tag{26}$$

The above space-time become special class of FRW K=0 model. It follows from [3] that proper affine vector fields in this case are given in equation (22). Cases (Q13) to (Q14) are explicitly the same.

**Case R1**

Here, we have $A_t(t,r,\theta) \neq 0$, $A_r(t,r,\theta) \neq 0$, $A_\theta(t,r,\theta) = 0$, $B_t(t,r,\theta) = 0$, $B_r(t,r,\theta) = 0$, $B_\theta(t,r,\theta) = 0$, $A_r^2(t,r,\theta) + 2A_{rr}(t,r,\theta) \neq 0$ and the rank of the $6\times 6$ Riemann matrix is one. There exist two linearly independent solutions $\theta_a = \theta_{,a}$ and $\phi_a = \phi_{,a}$ of equation (4) satisfying $\theta_{a;b} = 0$ and $\phi_{a;b} = 0$. From the above constraints, we get $A = A(t,r)$ and $B = m$, where $m \in R$. The line element after rescaling $\theta$ and $\phi$, can be written as

$$ds^2 = (-e^{A(t,r)} dt^2 + e^m dr^2) + d\theta^2 + d\phi^2. \tag{27}$$

The above space-time (27) is clearly 1+1+2 decomposable. Affine vector fields in this case are [4]

$$X = (\theta c_1 + \phi c_2 + c_3)\frac{\partial}{\partial\theta} + (\theta c_4 + \phi c_5 + c_6)\frac{\partial}{\partial\phi} + Y, \tag{28}$$

where $c_1, c_2, c_3, c_4, c_5, c_6 \in R$ and $Y$ is a homothetic vector field on each of two dimensional submanifolds of constant $\theta$ and $\phi$. The next step is to work out the homothetic vector field in the induced geometry of the submanifolds of constant $\theta$ and $\phi$. The non-zero components of the induced metric on each of the two-dimensional submanifolds of constant $\theta$ and $\phi$ are given by

$$g_{00} = -e^{A(t,r)} \quad g_{11} = e^m. \tag{29}$$

A vector field $Y$ is called a homothetic vector field if it satisfies equation (13). Expanding equation (13) in two dimensions and using (20) we get

$$A_t(t,r)X^0 + A_r(t,r)X^1 + 2\ X^0_{,0} = 2c, \tag{30}$$

$$e^m X^1_{,0} - e^{A(t,r)} X^0_{,1} = 0, \tag{31}$$

$$X^1_{,1} = c. \tag{32}$$

Solving the above equations we find the trivial solution of the above system which is

$$X^0 = X^1 = 0. \tag{33}$$

Affine vector fields in this case are

$$X = (0,0,\theta c_1 + \phi c_2 + c_3, \theta c_4 + \phi c_5 + c_6). \tag{34}$$

One can write the above equation (34) after subtracting the Killing vector fields as

$$X = (0,0,\theta c_1 + \phi c_2,\ \theta c_4 + \phi c_5). \tag{35}$$

Clearly in this case the above space-times (27) admit proper affine vector fields. Proper affine vector fields in this case are $\theta\theta_a$, $\phi\theta_a$, $\theta\phi_a$ and $\phi\phi_a$. Case (R3) is exactly the same.

**Case R3**

In this case $A_t(t,r,\theta) \neq 0$, $A_r(t,r,\theta) = 0$, $A_\theta(t,r,\theta) = 0$, $B_t(t,r,\theta) = 0$, $B_r(t,r,\theta) = 0$, $B_\theta(t,r,\theta) \neq 0$, $B_{\theta\theta}(t,r,\theta) = 0$ and the rank of the $6\times 6$ Riemann matrix is one. From the above constraints, we have $A = A(t)$ and $B = a\theta + b$, where $a,b \in R (a \neq 0)$. Here, there exist two linearly independent solutions $t_a = t_{,a}$ and $\theta_a = \theta_{,a}$ of equation (4). The vector field $t_a$ is covariantly constant whereas $\theta_a$ is not covariantly constant. The line element, after rescaling of $t$ can be written as

$$ds^2 = -\, dt^2 + e^{a\theta + b}(dr^2 + d\theta^2 + d\phi^2). \tag{36}$$

The above space-time (36) is clearly 1+3 decomposable. Substituting the above information in the affine equations, one finds that the affine vector fields in this case are

$$X^0 = c_4 t + c_5\theta + c_6,\quad X^1 = c_1\phi + c_2,\quad X^2 = c_7 t + c_8\theta + c_9,\quad X^3 = -\,c_1 r + c_3, \tag{37}$$

where $c_1, c_2, c_3, c_4, c_5, c_6, c_7, c_8, c_9 \in R$. One can write the above equation (37) after subtracting the Killing vector fields as

$$X = (c_4 t + c_5\theta,\ 0, c_7 t + c_8\theta + c_9,\ 0). \qquad (38)$$

In this case the above space-times (36) admit proper affine vector fields. Proper affine vector fields in this case are $tt_a$, $\theta t_a$, $t\theta_a$, $\theta\theta_a$ and $\theta_a$. Cases (R4) to (R6) are explicitly the same.

## Summary

In this paper a study of special non-static axially symmetric space times according to their proper affine vector fields is given. An approach is adopted to study proper affine vector fields for the above space-times using holonomy and decomposability, the rank of the $6\times 6$ Riemann matrix and direct integration. From the above discussion we obtain the following results:

(i) The case when the rank of the $6\times 6$ Riemann matrix is three there exists no non-trivial solution of equation (4). This is the space-time (7) and it admits an affine vector field which is Killing vector field (see case P1).

(ii) The case when the rank of the $6\times 6$ Riemann matrix is three there exists a unique nowhere zero independent timelike vector field which is a solution of equation (4) and is covariantly constant. This is the space-time (9) and it admits proper affine vector fields (see case Q1).

(iii) The case when the rank of the $6\times 6$ Riemann matrix is three there exists a unique nowhere zero independent timelike vector field which is a solution of equation (4) and is not covariantly constant. This is the space-time (26) and it admits proper affine vector fields (for details see case Q11).

(iv) The case when the rank of the $6\times 6$ Riemann matrix is one there exists two nowhere zero independent vector fields which are solutions of equation (4) and both are covariantly constant. This is the space-time (27) and it admits proper affine vector fields (see case R1).

(v) The case when the rank of the $6\times 6$ Riemann matrix is one there exists two nowhere zero independent vector fields which are solutions of equation (4) but only one is covariantly constant. This is the space–time (36) and it admits proper affine vector fields (see case R3).

It is important to mention here that the above techniques can also be used to study affine vector field in $f(R)$ gravity and the extended theories of gravity [15,16] in four dimensions.

## Acknowledgements

KSM thanks the NRF of South Africa for an innovation postdoctoral fellowship award for 2015.